\documentstyle[11pt,bezier,epsf,amsfonts]{article}

\begin{document}

\title{{\bf Kochen--Specker
Theorem: Two Geometric Proofs}\thanks{This paper has been completed
during the visits of the first author at the University of Technology
Vienna (1997) and of the third   author at the University of Auckland (1997).
The first   author has been
partially supported by AURC A18/XXXXX/62090/F3414056, 1996. The
second author was supported by DFG Research Grant No. HE 2489/2-1.}}
\author{{\large {\bf Cristian S.\  Calude},}\thanks{Computer Science Department,
The University of Auckland, Private Bag 92019, Auckland, New Zealand,
 e-mail: cristian@cs.auckland.ac.nz.} \quad
{\large {\bf Peter H.\ Hertling},}\thanks{Computer Science Department,
The University of Auckland, Private Bag 92019, Auckland, New Zealand,
e-mail: hertling@cs.auckland.ac.nz.}  \quad
{\large {\bf Karl  Svozil}}\thanks{Institut f\"ur Theoretische Physik,
University of Technology Vienna,
Wiedner Hauptstra\ss e 8-10/136,  A-1040 Vienna, Austria,
e-mail: svozil@tph.tuwien.ac.at.}}
\date{ }
\maketitle

\thispagestyle{empty}

\begin{abstract}
We present two geometric proofs for Kochen–Specker's theorem [S. Kochen, E. P. Specker: The problem of hidden variables in quantum mechanics, J. Math. Mech. 17 (1967), 59–87]. A quite similar argument has been used by Cooke, Keane, Moran [R. Cooke, M. Keane, W. Moran: An elementary proof of Gleason's theorem, Math. Proc. Camb. Phil. Soc. 98 (1985), 117–128], and by Kalmbach in her book to derive Gleason's theorem.
\end{abstract}

\section{Introduction}

The Kochen and Specker theorem \cite{kochen1}
(cf. also Specker \cite{specker-60}, Zierler and
 Schlessinger \cite{ZirlSchl-65} and John Bell \cite{bell-66}; see the
reviews by
Peres \cite{peres-91,peres}, Redhead \cite{redhead},
Clifton \cite{clifton-93}, Mermin \cite{mermin-93},
and Svozil and Tkadlec \cite{svozil-tkadlec}, among others)
-- as it is commonly argued, e.g.\ by
Peres \cite{peres} and Mermin \cite{mermin-93} -- is directed against
the noncontextual hidden parameter program envisaged by
Einstein, Podolsky and Rosen (EPR) \cite{epr}.
Indeed, if one
takes into account the entire Hilbert logic  (of dimension
larger than two) and if one considers all states thereon, any truth
value assignment to quantum propositions prior to the actual measurement
yields a contradiction. This can be proven by finitistic means, that is,
by considering only a finite number of one-dimensional closed linear
subspaces.

But, the Kochen Specker argument continues, it is always possible to prove
the existence of separable truth assignments for
classical propositional systems identifiable with
Boolean algebras. Hence, there does not exist any  injective
morphism from a quantum logic into some Boolean algebra.

Rather than rephrasing the Kochen and Specker argument
\cite{kochen1} concerning nonexistence
of truth assignments in three-dimensional Hilbert logics in its original form
or in terms of less
subspaces (cf.\ Peres \cite{peres}, Mermin \cite{mermin-93}), or of Greechie
diagrams, which represent co--measurability (commutativity) very nicely
(cf.\ Svozil and Tkadlec \cite{svozil-tkadlec},
Svozil \cite{svozil-ql}), we shall give two geometric
arguments which are derived from proof methods for Gleason's theorem
(see Piron \cite{piron-76}, Cooke, Keane, and Moran \cite{c-k-m},
and Kalmbach \cite{kalmbach-86}).

Let $L$ be the lattice of closed linear subspaces of the
three-dimensional real Hilbert space ${\Bbb R}^3$. A {\em two-valued
probability measure} on $L$ is a map
$v:L\to\{0,1\}$ which maps the zero-dimensional subspace
containing only the origin $(0,0,0)$ to $0$, the full space
${\Bbb R}^3$ to $1$, and which is additive on orthogonal subspaces.
This means that for two orthogonal subspaces $s_1, s_2 \in L$
the sum of the values $v(s_1)$ and $v(s_2)$ is equal to the
value of the linear span of $s_1$ and $s_2$. Hence,
if $s_1, s_2, s_3 \in L$ are a tripod of pairwise orthogonal
one-dimensional subspaces, then
\[ v(s_1) + v(s_2) + v(s_3) = v({\Bbb R}^3) = 1. \]
The two-valued probability measure $v$ must map one of these subspaces
to $1$ and the other two to $0$.
We will show that there is {\it no}  such map.
In fact, we will show the assertion $(*)$:
\begin{quote}
{\em there is no map $v$ which is defined on all
one-dimensional subspaces of ${\Bbb R}^3$ and maps
   exactly one subspace out of each tripod of pairwise
      orthogonal one-dimensional subspaces to $1$ and the other two to $0$}.
\end{quote}

In the following  we often identify
a one-dimensional subspace of ${\Bbb R}^3$ with one of its two intersection
points with the unit sphere
\[ S^2 = \{x \in {\Bbb R}^3 \ | \ ||x||=1\}\,. \]
In the statements  ``a point (on the unit sphere) has value $0$
(or value $1$)'' or
that ``two points (on the unit sphere) are orthogonal'' we always
mean the corresponding one-dimensional subspaces.
Note also that the intersection of a two-dimensional subspace with
the unit sphere is a great circle.

\section{First proof}

To start  the first proof,
let us assume that a function $v$ satisfying the above condition
exists.
Let us consider an arbitrary tripod of orthogonal points
and let us fix the point with value $1$. By a rotation we
can assume that it is the north pole with the coordinates
$(0,0,1)$. Then, by the condition above, all points
on the equator $\{(x,y,z) \in S^2\ | \ z=0\}$ must have value $0$
since they are orthogonal to the north pole.

\begin{figure}[htbp]
\centerline{ \epsfxsize12cm \epsfbox{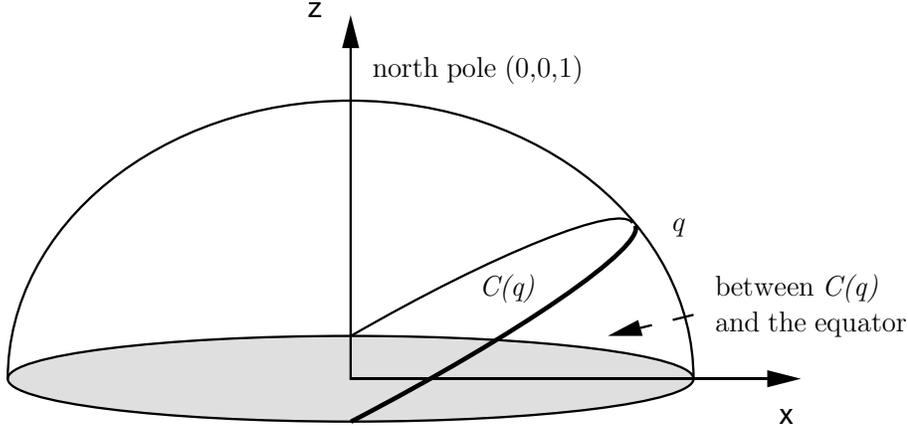} }
\vskip-2cm
\caption{The great circle $C(q)$}
\label{figure:greatcircle}
\end{figure}

Let $q=(q_x,q_y,q_z)$ be a point in the northern
hemisphere, but not equal to the north pole, that is
$0< q_z < 1$. Let $C(q)$ be the unique
great circle which contains $q$ and the points
$\pm(q_y,-q_x,0)/\sqrt{q_x^2+q_y^2}$ in the equator, which are orthogonal
to $q$.
Obviously,  $q$ is the northern-most point on $C(q)$.
To see this, rotate the sphere around the $z$-axis
so that $q$ comes to lie in the $\{y=0\}$-plane;
see Figure \ref{figure:greatcircle}.
Then the two points in the equator orthogonal to $q$ are
just the points $\pm(0,1,0)$, and $C(q)$ is the intersection
of the plane through $q$ and $(0,1,0)$ with the unit sphere, hence
\[C(q) = \{p \in {\Bbb R}^3 \ | \ (\exists \ \alpha,\beta \in {\Bbb R}) \
     \alpha^2 + \beta^2 =1 \ \mbox{\rm and } p=\alpha q + \beta
     (0,1,0) \}\,.\]
\newpage
This shows that $q$ has the largest $z$-coordinate among all
points in $C(q)$.

\bigskip

\bigskip

Assume that $q$ has value $0$. We claim that then all points
on $C(q)$ must have value $0$. Indeed, since $q$ has value
$0$ and the orthogonal point $(q_y,-q_x,0)/\sqrt{q_x^2+q_y^2}$
on the equator also has value $0$, the one-dimensional subspace
orthogonal to both of them must have value $1$.
But this subspace is orthogonal to all points on $C(q)$.
Hence all points on $C(q)$ must have value $0$.

Now, still assuming that $q$ has value $0$,
we consider an arbitrary point $\tilde{q}$ on
$C(q)$ in the northern hemisphere. We have just seen
that $\tilde{q}$ has value $0$.
We claim that now by the same argument as above also all points on the
great circle $C(\tilde{q})$ must have value $0$.
Namely, $C(\tilde{q})$ is the unique
great circle which contains $\tilde{q}$ and the points
$\pm(\tilde{q}_y,-\tilde{q}_x,0)/\sqrt{\tilde{q}_x^2+\tilde{q}_y^2}$
in the equator, which are orthogonal to $\tilde{q}$.
Since $\tilde{q}$ has value $0$ and the orthogonal point
$(\tilde{q}_y,-\tilde{q}_x,0)/\sqrt{\tilde{q}_x^2+\tilde{q}_y^2}$
in the equator has value $0$, the one-dimensional subspace orthogonal
to both of them must have value $1$.
But this subspace is orthogonal to all points on $C(\tilde{q})$.
Hence all points on $C(\tilde{q})$ must have value $0$.

The great circle $C(q)$ divides the northern hemisphere into two
regions, one containing the north pole,
the other consisting of the points below $C(q)$ or
``lying between $C(q)$ and the equator'',
see Figure \ref{figure:greatcircle}.
The circles $C(\tilde{q})$ with $\tilde{q} \in C(q)$
certainly cover the region between $C(q)$ and the equator.\footnote{This will
be shown formally in the proof of the geometric lemma below.}
Hence any point in this region must have value $0$.

But the circles $C(\tilde{q})$ cover also a part
of the other region. In fact, we can iterate this process.
We say that a point $p$ in the northern hemisphere
{\em can be reached} from a point $q$ in the
northern hemisphere, if there is a finite
sequence of points $q=q_0, q_1, \ldots, q_{n-1}, q_n=p$
in the northern hemisphere such that $q_i\in C(q_{i-1})$
for $i=1,\ldots,n$.
Our consideration above shows that if $q$ has value $0$ and
$p$ can be reached from $q$, then also $p$ has value $0$.

The following geometric lemma due to Piron \cite{piron-76}
(see also
Cooke, Keane, and Moran \cite{c-k-m} or Kalmbach \cite{kalmbach-86})
is a consequence of the fact that the curve $C(q)$ is tangent to
the horizontal plane through the point $q$.
\begin{quote}
{\it If $q$ and $p$ are points in the northern hemisphere
with $p_z < q_z$, then $p$ can be reached from $q$.}
\end{quote}
This lemma will be proved in an appendix.
We conclude that, if a point $q$ in the northern hemisphere has value $0$,
then every point $p$ in the northern hemisphere with $p_z < q_z$ must
have value $0$ as well.

Consider the tripod $(1,0,0), (0,{1 \over \sqrt{2}},{1 \over \sqrt{2}}),
(0,-{1 \over \sqrt{2}},{1 \over \sqrt{2}})$. Since $(1,0,0)$ (on the equator)
has value $0$,
one of the two other points has value $0$ and one has
value $1$. By the geometric lemma and our above considerations
this implies that all points $p$ in the northern
hemisphere with $p_z<{ 1\over\sqrt{2}}$ must have value $0$
and all points $p$ with $p_z>{1\over\sqrt{2}}$ must have value $1$.
But now we can choose any point $p^\prime$ with
${1\over\sqrt{2}} < p^\prime_z < 1$
as our new north pole and  deduce that the function $v$
must have the same form with respect to this
pole. This is clearly impossible.
Hence, we have proved our assertion $(*)$.

\section{Second  proof}

In the following we give a second topological and
geometric proof for $(*)$.
In this proof we shall not use the geometric lemma above.

Fix an arbitrary point on the unit sphere with value $0$.
The  great circle consisting of points orthogonal to this
point splits into two disjoint sets, the set of points
with value $1$, and the set of points orthogonal to these
points. They have value $0$. If one of these two
sets were open, then the other had to be open as well.
But this is impossible since the circle is connected
and cannot be the union of two disjoint open sets.
Hence the circle must contain a point $p$ with value $1$
and a sequence of points $q(n)$, $n=1,2,\ldots$ with value $0$ converging
to $p$. By a rotation we can assume that $p$ is the
north pole and the circle lies in the $\{y=0\}$-plane.
Furthermore we can assume that all points $q_n$ have
the same sign in the $x$-coordinate. Otherwise,  choose
an infinite subsequence of the sequence $q(n)$ with this property.
In fact, by a rotation we can assume that all points $q(n)$ have
positive $x$-coordinate (i.e.\ all points $q(n)$, $n=1,2,\ldots$
lie as the point $q$ in  Figure \ref{figure:greatcircle} and approach the
northpole as $n$ tends to infinity).
All points on the equator have value $0$.
By the first step in the proof of the geometric lemma in the appendix,
all points in the northern hemisphere which lie between $C(q(n))$
(the great circle through $q(n)$ and $\pm(0,1,0)$)
and the equator can be reached from $q(n)$. Hence, as we
have seen in the first proof, $v(q(n))=0$ implies that
all these points must have value zero. Since $q(n)$ approaches
the northpole, the union of the regions between $C(q(n))$ and
the equator is equal to the open right half
$\{q \in S^2\ | \ q_z>0, q_x>0\}$
of the northern hemisphere.
%
%
 Hence all points in this set
have value $0$.  Let $q$ be a point in the left half
$\{q \in S^2\ | \ q_z>0, q_x<0\}$ of the northern hemisphere.
It forms a tripod together with the point
$(q_y,-q_x,0)/\sqrt{q_x^2+q_y^2}$ in the equator and the point
\[(-q_x,-q_y,{q_x^2+q_y^2 \over q_z}) /
        ||(-q_x,-q_y,{q_x^2+q_y^2 \over q_z})||\]
in the right half. Since these two points have value $0$,
the point $q$ must have value $1$. Hence all points
in the left half of the northern hemisphere must have
value $1$. But this leads to a contradiction because
there are tripods with two points in the left half,
for example the tripod
$(-{1 \over 2},{1 \over \sqrt{2}},{1 \over 2})$,
$(-{1 \over 2},-{1 \over \sqrt{2}},{1 \over 2})$,
$({1 \over \sqrt{2}},0,{1 \over \sqrt{2}})$.
This completes the second proof for $(*)$ and, hence, for
the fact that there is no two-valued probability measure
on the lattice of subspaces of the three-dimensional
Euclidean space which preserves the lattice operations
at least for orthogonal elements.

\section{Final comments}
Do the partial order and  lattice operations of a quantum logic correspond
to the logical implication and connectives of classical logic?
 Kochen and Specker's theorem answers the above question in the negative.
However, this answer is just one among different possible ones, not all
negative.
In a forthcoming article
\cite{CalHerSvo}
we discuss the above question in terms of mappings of quantum
worlds into classical ones, more specifically, in terms of embeddings of
quantum logics
into classical logics; depending upon the type of restrictions imposed on
embeddings the question may get negative or positive answers.

\appendix
\section*{Appendix: Proof of the geometric lemma}

\begin{figure}[htbp]
\setlength{\unitlength}{1cm}
\begin{center}
\begin{picture}(8,5)
   \thinlines
   \put(2,2.5){\circle*{0.2}}
   \put(5,2.5){\circle*{0.2}}
   \put(4.7,2.8){\line(1,0){0.3}}
   \put(4.7,2.5){\line(0,1){0.3}}
   \multiput(2,2.5)(0.2,0){15}{\line(1,0){0.1}}
   \put(5,0){\line(0,1){5}}
   \put(2,3.5){\makebox(0,0){the north pole}}
   \put(2,3){\makebox(0,0){$(0,0,1)$}}
   \put(6.5,3.5){\makebox(0,0)[l]{the image of the}}
   \put(6.5,3){\makebox(0,0)[l]{region between $C(q)$}}
   \put(6.5,2.5){\makebox(0,0)[l]{and the equator}}
   \put(4.5,2.1){\makebox(0,0){$h(q)$}}
   \put(5.9,0.5){\makebox(0,0){$h(C(q))$}}
\end{picture}
\end{center}
\caption{The plane $H$ viewed from above}
\label{figure:projectionh}
\end{figure}
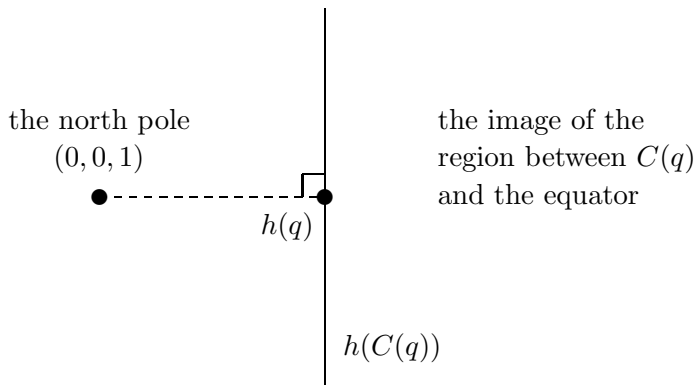

In this appendix we are going to prove the geometric lemma
due to Piron \cite{piron-76} which was formulated in Section 2.2.
First let us restate it. Consider a point $q$ in the northern hemisphere
of the unit sphere $S^2 = \{p \in {\Bbb R}^3 \ | \ ||p||=1\}$. By
$C(q)$ we denote the unique
great circle which contains $q$ and the points
$\pm(q_y,-q_x,0)/\sqrt{q_x^2+q_y^2}$ in the equator, which are orthogonal
to $q$, compare Figure \ref{figure:greatcircle}.
We say that a point $p$ in the northern hemisphere {\em can be reached}
from a point $q$ in the
northern hemisphere, if there is a finite
sequence of points $q=q_0, q_1, \ldots, q_{n-1}, q_n=p$
in the northern hemisphere such that $q_i\in C(q_{i-1})$
for $i=1,\ldots,n$. The lemma states:
\begin{quote}
{\it If $q$ and $p$ are points in the northern hemisphere
with $p_z < q_z$, then $p$ can be reached from $q$.}
\end{quote}
For the proof we follow
Cooke, Keane, and Moran \cite{c-k-m} and Kalmbach \cite{kalmbach-86}).
We consider the
tangent plane $H=\{p \in {\Bbb R}^3\ | \ p_z=1\}$ of the unit sphere
in the north pole  and the projection $h$ from the northern hemisphere
onto this plane which maps each point
$q$ in the northern hemisphere to the intersection $h(q)$
of the line through the origin and $q$ with the plane $H$.
This map $h$ is a bijection. The north pole $(0,0,1)$ is mapped
to itself. For each $q$
in the northern hemisphere (not equal to the north pole)
the image $h(C(q))$ of the great
circle $C(q)$ is the line in $H$ which goes through $h(q)$
and is orthogonal to the line through the north pole  and
through $h(q)$.
Note that $C(q)$ is the intersection of a plane with $S^2$, and
$h(C(q))$ is the intersection of the same plane with $H$;
see Figure \ref{figure:projectionh}.
The line $h(C(q))$ divides
$H$ into two half planes. The half plane not containing the north pole
is the image of the region in the northern
hemisphere between $C(q)$ and the equator.
Furthermore note that $q_z > p_z$ for two points in
the northern hemisphere if and only if $h(p)$ is further
away from the north pole than $h(q)$.
We proceed in two steps.

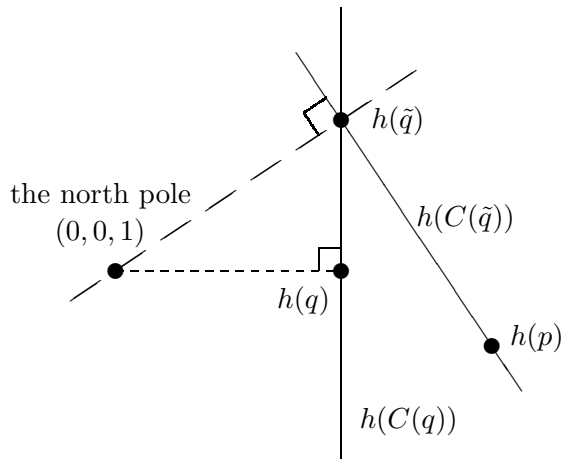
\begin{figure}[htbp]
\setlength{\unitlength}{1cm}
\begin{center}
\begin{picture}(9,6)
   \thinlines
   \put(2,2.5){\circle*{0.2}}
   \put(5,2.5){\circle*{0.2}}
   \put(4.7,2.8){\line(1,0){0.3}}
   \put(4.7,2.5){\line(0,1){0.3}}
   \multiput(2,2.5)(0.2,0){15}{\line(1,0){0.1}}
   \put(5,0){\line(0,1){6}}
   \put(1.8,3.5){\makebox(0,0){the north pole}}
   \put(1.8,3){\makebox(0,0){$(0,0,1)$}}
   \multiput(1.4,2.1)(0.6,0.4){8}{\line(3,2){0.4}}
   \put(5,4.5){\circle*{0.2}}
   \bezier{40}(4.7,4.3)(4.6,4.45)(4.5,4.6)
   \bezier{40}(4.5,4.6)(4.65,4.7)(4.8,4.8)
   \put(4.4,5.4){\line(2,-3){3}}
   \put(7,1.5){\circle*{0.2}}
   \put(4.5,2.1){\makebox(0,0){$h(q)$}}
   \put(7.6,1.6){\makebox(0,0){$h(p)$}}
   \put(5.4,4.4){\makebox{$h(\tilde{q})$}}
   \put(6,3.1){\makebox{$h(C(\tilde{q}))$}}
   \put(5.9,0.5){\makebox(0,0){$h(C(q))$}}
\end{picture}
\end{center}
\caption{The point $p$ can be reached from $q$}
\label{figure:belowcq}
\end{figure}

Step 1. First, we show that, if $p$ and $q$ are points in the northern
hemisphere and $p$ lies in the region between $C(q)$ and the
equator, then $p$ can be reached from $q$. In fact, we show that
there is a point $\tilde{q}$ on $C(q)$ such that $p$ lies on $C(\tilde{q})$.
Therefore we consider the images of $q$ and $p$ in the plane $H$;
see Figure \ref{figure:belowcq}.
The point $h(p)$ lies in the half plane bounded by $h(C(q))$ not containing
the north pole.
Among all points $h(q^\prime)$ on the line $h(C(q))$ we set $\tilde{q}$
to be one of the two points such that the line trough the north pole and
$h(q^\prime)$
and the line through $h(q^\prime)$ and $h(p)$ are orthogonal. Then this last
line is the image of $C(\tilde{q})$, and $C(\tilde{q})$ contains the point $p$.
Hence $p$ can be reached from $q$. Our first claim is proved.

\begin{figure}[htbp]
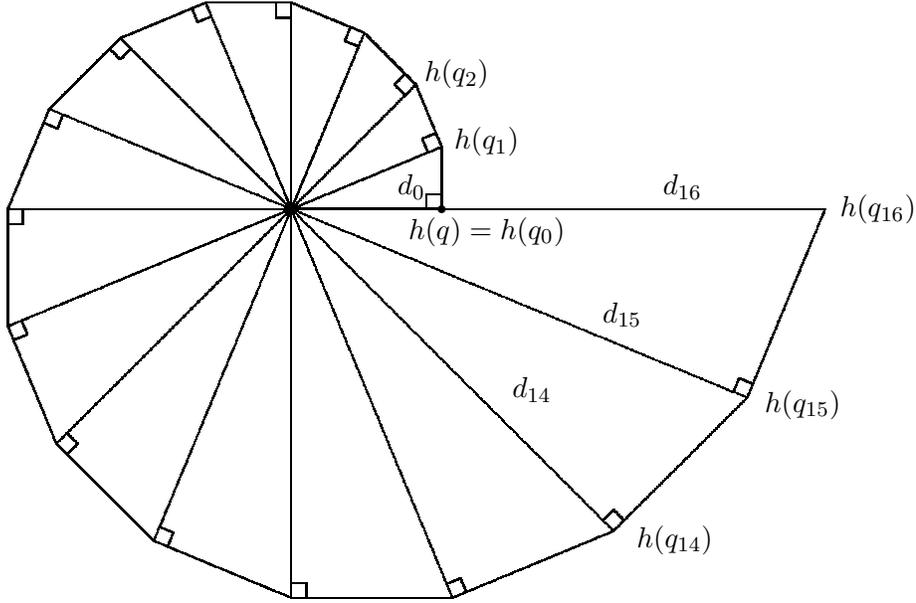

\setlength{\unitlength}{2cm}
\begin{center}
\input shell.eps
\end{center}
\caption{The shell in the plane $H$ for $n=16$}
\label{figure:schnecke}
\end{figure}

Step 2. Fix a point $q$ in the northern hemisphere.
Starting from $q$ we can wander around the northern hemisphere along
great circles of the form $C(p)$ for points $p$ in the following way:
for $n\geq 5$ we define a sequence $q_0, q_1, \ldots, q_n$
by setting $q_0=q$ and by choosing $q_{i+1}$ to be that point on
the great circle $C(q_i)$ such
that the angle between $h(q_{i+1})$ and $h(q_i)$ is
$2\pi/n$. The image in $H$ of this configuration is a
shell where $h(q_n)$ is the point furthest away
from the north pole;  see Figure \ref{figure:schnecke}.
First, we claim that any point $p$ on the unit sphere with
$p_z < (q_n)_z$ can be reached from $q$. Indeed, such a point
corresponds to a point $h(p)$ which is further away from the
north pole than $h(q_n)$.
There is an index $i$
such that $h(p)$ lies in the half plane bounded by  $h(C(q_i))$
and not containing the north pole, hence such that
$p$ lies in the region between $C({q_i})$ and the equator.
Then, as we have already seen, $p$ can be reached from
$q_i$ and hence also from $q$.
Secondly, we claim that $q_n$ approaches
$q$ as $n$ tends to infinity. This is equivalent to
showing that the distance of $h(q_n)$ from $(0,0,1)$
approaches the distance of $h(q)$ from $(0,0,1)$.
Let $d_i$ denote the distance of $h(q_i)$ from $(0,0,1)$
for $i=0,\ldots,n$. Then
$d_i / d_{i+1} = \cos(2\pi/n)$, see Figure
\ref{figure:schnecke}. Hence
$d_n = d_0 \cdot \cos(2\pi/n)^{-n}$.
That $d_n$ approaches $d_0$ as $n$ tends to infinity
follows immediately from the fact that
$\cos(2\pi/n)^n$ approaches $1$ as $n$ tends to infinity.
 For
completeness sake\footnote{Actually, this is an exercise in elementary
analysis.} we prove it by
proving the equivalent statement that $\log(\cos(2\pi/n)^n)$ tends to $0$
as $n$ tends to
infinity. Namely, for small $x$ we know the formulae
$\cos(x)=1-x^2/2 + {\cal O}(x^4)$ and
$\log(1+x)=x+{\cal O}(x^2)$.
Hence, for large $n$,
\begin{eqnarray*}
   \log(\cos(2\pi/n)^n) & = &
      n \cdot \log(1-2{\pi^2 \over n^2} + {\cal O}(n^{-4})) \\
   & = & n \cdot ( - 2 {\pi^2 \over n^2} + {\cal O}(n^{-4}))\\
       & = & - {2 \pi^2 \over n} + {\cal O}(n^{-3}) \, .
\end{eqnarray*}
This ends the proof of the geometric lemma.

\section*{Acknowledgement} In writing the final version of this paper we
benefitted from Anatolij Dvure{\v{c}}enskij's critical remarks.


\end{document}